\documentstyle[12pt,aasms4]{article}
\input epsf
\def\diff{\rm d}
\def\wig#1{\mathrel{\hbox{\hbox to 0pt{%
\lower.5ex\hbox{$\sim$}\hss}\raise.4ex\hbox{$#1$}}}}

\begin{document}

\title{\bf The effect of electron gas polarization on thermonuclear reaction rates
   in dense stars}      

 \author{{\sc Mikael Sahrling$^1$ and Gilles Chabrier$^2$}
\\ $^1$California Institute of Technology, 130-33, Pasadena, CA
 91125
\\ $^2$Centre de Recherche Astrophysique de Lyon (UMR CNRS 5574),
\\ Ecole Normale Sup\'erieure, 69364 Lyon Cedex 07, France\\}

\begin{abstract} In dense stars the nuclear reaction rates are  
  influenced by screening effects arising from both ions and electrons. In this paper we calculate the enhancement factors due to electron 
polarization in the high-density, degenerate and relativistic
regime, for 
non-resonant
nuclear reaction rates. In an earlier analysis, Sahrling had proposed the 
possibility that the polarized electrons would lower the reaction rate 
instead of enhancing it. This analysis was based on Monte Carlo simulations with only one choice of density, temperature and charge. Here we extend the analysis to a wider range of
densities, temperatures and charges and calculate analytical expressions for
the enhancement factors. We concentrate on carbon and oxygen ions and
show that at very high-densities,
high-order quantum effects will be important and act to reduce the zeroth order,
classical value for the enhancement factor. We show that in any case, the total electron contribution remains weak,
namely an enhancement in the reaction of about a 
factor 2, contrarily to what had been claimed by some authors in previous
calculations.
We examine the astrophysical implications of these results on the final
stages of massive white dwarfs, near the carbon-ignition curve.

\bigskip \bigskip

Subject headings : dense matter --- nuclear reactions ---
stars: interiors, white dwarfs --- stars : supernovae

\end{abstract}

\section{Introduction}
The rate of nuclear reactions plays an important role in many fields of
astrophysics. It sets the time scale for various processes, such as energy 
release in
the core of stars. It is also a crucial factor to determine the fate 
of accreting
white dwarfs in binary systems (see e.g. Isern \& Hernanz 1994). 
In dense matter, the surrounding ions screen the Coulomb barrier between the two
reacting nuclei and thus increase the reaction rate compared to the infinitely 
dilute plasma,
as shown originally by Schatzman (1948). One usually assumes that
the electrons can be treated as a uniform, {\it rigid} background, the so-called one-component-plasma (OCP) or binary ionic mixture (BIM) model. This approximation has
been studied extensively by many authors and for recent discussions see e.g.
Yakovlev \& Shalybkov (1989), Sahrling (1994b).
In general, electron polarization effects are small since the
Thomas-Fermi screening length is larger than the mean distance
between the ions. Moreover people have
generally concentrated on the so-called classical or zeroth order contribution to the reaction
rate (see for example, Mochkovitch \& Hernanz 1986, Ichimaru \& Utsumi 1983, Yakovlev \& Shalybkov 1989,
Ichimaru and Ogata 1991, Sahrling 1994 b), although the quantum mechanical, or higher order, contributions are discussed to some extent in these papers. Sahrling (1994b)
found that the quantum corrections give a much smaller contribution to the
reaction rate than suggested by some previous calculations. He also 
proposed that the electrons would reduce the reaction rate and this could affect the 
late stages of stellar evolution. The investigation was
based on a mean (or screening-) potential that was calculated for only
one choice of density, temperature and charge for the ionic plasma.
In this paper, we examine in details the effect of electron gas polarization
in the high-density regime corresponding to degenerate,
relativistic electrons.
We extend the previous calculations to a wider range of densities, charges and
temperatures and we calculate the quantum correction to the classical
enhancement factor. 
\newline\indent
In Sect. 2 we describe how the 
polarization of the electron gas is taken into account and we derive analytical 
formulae for the corresponding correction to the reaction rate. The 
astrophysical consequences are described in Sect. 3 
followed by the conclusion in Sect. 4.

\section{Reaction rate due to electron gas polarization}
This section is divided into three subsections. In Sect. 2.1, we briefly review the 
theory of reaction rates in dense matter and we describe the calculations of
the correction arising from the 
electron gas polarization. We concentrate on densities and temperatures
where the electrons are degenerate and relativistic, since, as we argue below, we 
expect quantum effects to be most significant in that region. Section 2.2 describes the calculations of
the mean potential and 
analytical fitting formulae for these potentials are given at the end of the subsection. The enhancement factors due to a 
polarized electron gas are calculated in Sect. 2.3 and
analytical fitting formulae are also provided for these factors.
\subsection{Basic formalism}
When investigating non-resonant reaction rates in dense matter one usually discusses  
enhancement factors, where one normalises the reaction rate in the correlated
plasma to its value 
in an 
infinitely dilute gas. The advantage of this procedure is
 that the matrix element 
describing the actual nuclear reaction, which is considered to be 
independent of the neighbour ions and electrons, is cancelled out and one 
is left with a simpler expression (see e.g. Alastuey \& Jancovici 1978). For a recent discussion
 of some of the 
basic approximations and assumptions see Sahrling (1994a,b). One 
assumes usually that the electrons are uniformly and homogeneously
distributed in the plasma, the so-called rigid background approximation. This
yields the well-studied 
OCP model (or BIM for a binary mixture) to describe the thermodynamic properties of dense matter. In this paper, we examine the case where such an approximation is no longer
valid and where the {\it inhomogeneous} electron gas is polarized by the external field due to the 
surrounding nuclei. 
\newline\indent
The plasma is described by $N$ ions of mass $m$ and charge $Ze$ moving in a volume $V$. The density $\rho$ and 
temperature $T$ are such that that the electrons are degenerate and completely stripped off the atoms, so that the electroneutrality condition yields
$N_e=N\times Z$, where $N_e$ is the number of free electrons.
 We restrict the present study to the region below the neutronisation threshold. For ions such as $C^{6+}$ and $O^{8+}$ 
all these conditions correspond to $10^5~{\rm g~cm^{-3}} < \rho < 10^{10}~{\rm g~cm^{-3}}$ and 
 $10^7~{\rm K} < T < 10^9~{\rm K}$. The physical state of the one-component-plasma is described universally by two
parameters. The coupling parameter $\Gamma$ is simply the ratio of the mean Coulomb energy
to the average kinetic energy of the ions,
$$\Gamma= \beta{(Ze)^2\over{a}},\;\;\;\beta={1\over {k_{\rm B}T}} 
\;\eqno(1)$$
where $k_{\rm B}$ is the Boltzmann constant. The Wigner-Seitz radius $a$
  is defined as 
$a=({3 V/(4\pi
N)})^{1/3}$. Above
$\Gamma \simeq 172$, the crystal becomes stable and the matter is solid
(Dubin, 1990). Since we consider presently reaction rates in the fluid phase, we require $\Gamma < 172$. Relativistic effects on the electrons velocity are measured
by the parameter
$$ x = {p_{\rm F}\over {m_{\rm e} c}},\eqno(2)$$
where $p_{\rm F}$ is the Fermi-momentum of the electrons and $m_{\rm e}$ denotes
their mass.
To measure quantum effects we use the 
parameter
$\delta$, which is defined as 
$$\delta={b_0\over a}=\left({16\hbar^2\beta^2 (Ze)^2 N\over{3\pi m V}}\right)^{1\over 3} \sim x^{1/3}\,\Gamma^{2/3} \sim 
\left({Z\over A}\right)^{2\over 3} T^{-{2\over 3}} \rho^{1\over 3}\;,\eqno(3)$$
where $b_0$ is the classical turning point at the Gamow peak incident 
energy for the infinitely dilute plasma. The parameter $\delta$ corresponds to the more widely used combination $\delta=3\Gamma/\tau$ where $\tau$ is the penetration probability of the Coulomb barrier for the infinitely dilute plasma.
Quantum effects become increasingly important as 
$\delta$ gets larger, i.e. for small temperatures and high densities. In this paper we 
will examine densities of the carbon/oxygen plasma where the electrons are highly relativistic ($x>>1$), as commonly encountered in the core of massive white dwarfs.
However, in order to avoid exchange effects between the {\it ions}, we will restrict the calculations to the region in the ($\rho,T$)-plane where the 
ionic thermal de 
Broglie
wavelength is smaller than the mean inter-ionic distance, i.e. $\lambda=(2\pi\hbar^2/mk_{\rm B}T)^{1/2}\ll a$.
 In terms of $\Gamma$ and
$\delta$
this condition can be written, using $\lambda/a\simeq 3.9
\sqrt{\delta^3/\Gamma}$,
$$\delta \, < \, 0.4\, \Gamma^{1/3}\;\eqno(4)$$
For carbon and oxygen ions under white dwarf central density conditions, $\delta\sim 2$ when $\Gamma\sim 180$ so the condition
(4) is valid below the line of solidification, i.e. in the entire fluid phase. The lowest order correction to the
final reaction rate at this limit,
in $\lambda/a$, is about 50\% (Sahrling 1994b).
\newline\indent The effective interionic 
potential $V^{\rm e}(\rho,r)$, which takes into account polarization effects of the electron gas, is
given by
$$V^{\rm e}(\rho, r) = {(Ze)^2\over {2\pi^2}}\int \diff{\vec k} 
{1\over{k^2\epsilon(\rho, k)}}
\exp(i{\vec k\cdot \vec r})\;, \eqno(5)$$

where $k=|{\vec k}|$ is the wavenumber (Hubbard \& Slattery 1971; Galam \& Hansen 1976; Yakovlev \& Shalybkov 1989;
Chabrier 1990).
For $\epsilon(k)$ we use the static dielectric function of 
relativistic
degenerate electrons (Jancovici 1962). We exclude
the vacuum polarization part of $\epsilon$ since that will not affect the 
rate by more
than a few percent (Gould 1990).
\newline\indent
 Let us now assume that
 two particles are 
moving towards each other. The probability of a nuclear reaction between
them is given 
by the pair distribution function $g(d)$ where $d$ is of nuclear 
dimensions. Since in coulombic matter $d$ is much smaller than $a$ one can evaluate the 
pair distribution at $d=0$ (Alastuey \& Jancovici 1978). If the nuclear reaction itself
 is 
independent of the neighbour
ions one can normalise the rate to the infinitely dilute plasma. 
The enhancement factor $E$ is then given by

$$E={g(0)\over g_0(0)}\;,\eqno(6)$$
where 
$$g(0)=g(r=0)=<{\vec 0}|\exp(-\beta H)|{\vec 0} >
\eqno(7)$$
and
$$H=-{\hbar^2\over {m}} \Delta + V^{\rm e}(\rho,r) + w(r)\;
\eqno(8)$$
$g_0(r)$ is the pair distribution function with $w=0$ and $r=|{\vec r}|$ is the distance between the
reacting particles. The potential felt
 by the reacting pair from the
neighbour ions is approximated by a mean potential
 $w(r)$, defined by:
$${\rm e}^{-\beta w(r)}=\int_{V^{N-1}}\int \exp[-\beta \bigl( 
W({\vec
r,\vec R,\alpha}
)- F\bigr) ] {\diff\alpha \diff{\vec R}\over V^{N-1}}\;.
\eqno(9)$$
where  $\alpha$ denotes 
the coordinates of all neighbour ions in the volume $V$ and  $\vec R$ is the
 center-of-mass coordinate of the reacting pair. $W({\vec
r,\vec R,\alpha}
)$ denotes the sum of all pair interactions except the one between the reacting pair.
$F$ is the Helmholtz free energy of the system. The accuracy
of the mean-potential approach has been explored in details by Jancovici (1977), Alastuey \& Jancovici 
(1978), J\"andel \& Sahrling (1992) and Sahrling (1994a).
\newline\indent
The right-hand-side of eqn. (7) denotes a
quantum mechanical matrix element which can be calculated, for instance, by the
path integral formalism (Feynman \& Hibbs 1965). Keeping only the classical 
action in the functional integral yields the following expression :
$$E={{\int_0^\infty \exp(-S(E')/\hbar) \diff E'}\over{\int_0^\infty
\exp(-S_0(E')/\hbar) \diff E'}}\left(1-{\cal O}\left[(\lambda/a)^2\right]\right)\eqno(10)$$
where
$$ S(E')/\hbar = {2\over{\hbar}}\int_0^b \{m[V^{\rm e}(\rho,r) + w(r) -
E']\}^{1/2}\diff r
+\beta E'\;.
\eqno(11)$$
The function $S_0$ in Eq. (10) is the same as $S$ but for the bare Coulomb potential $V(r)=(Ze)^2/r$ and $w=0$. The first term
in the integrand of $S$ in (11) is the penetration probability through the combined potential $V^{\rm e}(\rho,r)+
w(r)$ and the second term is simply the thermal weight. For details of the
derivation, see Alastuey \& Jancovici (1978) and Sahrling (1994a,b).
\subsection{Mean Potentials}
 We have calculated the mean 
potential in Eq. (9) for various combinations of the density and temperature
for carbon and oxygen ions. 
We employ the Monte Carlo so-called ``minimum-image convention"
(Brush, Sahlin \& Teller 1966). According to this scheme, a particle in the basic box is allowed to interact only
with each of the $N-1$ other particles in the basic box or with the nearest ``image" of this particle in one of the neighbouring cells. In other words, 
each particle interacts with the $N-1$ particles that happen to be located
in a cube centered at the particle at any time. This method is not useful for 
long-range interactions, such as the bare Coulomb interaction.
In our case the {\it screened} effective potential is of Yukawa-type, i.e. 
of short-range type. We found that we needed at least 2000
particles in the basic Monte Carlo box in order to avoid border effects for $0<r/a\le 2$. 
In general we require the accuracy in $w/\Gamma$ to be better than $10^{-3}$. This leads
to an uncertainty in $w$ less than 20 \%. In Fig. 1 we show the result
for $\Gamma=120$, $x=10$ and $Z=6,8$.
 \newline\indent
It is too time-consuming to use Eq. (5) explicitly 
to calculate the interaction
between  each pair of particles in the Monte Carlo simulation. We therefore calculate
$V^{\rm e}(\rho,r)$
at 10000 points between $[0,L]$ and use the result as an input vector to the Monte Carlo code where we interpolate
the vector linearly. The accuracy in $w/\Gamma$ of this procedure was found to be better than
 $10^{-3}$ by comparing with runs using 20000 points. 
 For each 
combination of density, temperature and charge we used $10^7$ 
configurations where we started from a random configuration. We ignored the initial $10^6$ configurations because they have not reached equilibrium. We also made
 sure that this number of configurations was adequate by comparing with runs using $2\times 10^7$ configurations.
The code we use is based on a code kindly supplied by W.L. 
Slattery for the OCP (Slattery et al. 1980).

The variation of the mean potential (9) with $x$ is smaller than 10 \% for
$x>10$. There is, however, a strong dependence on $Z$ which is consistent 
with the analysis of Yakovlev \& Shalybkov (1989).
We have constructed the following fitting formula for the screening potential :
$${\beta w(r)\over\Gamma}=\cases{&$-{h_0^{\rm e}\over\Gamma}+h_1^{\rm e} 
r^2+h_2^{\rm e}
r^4\,,\; r < r_0$\cr
&$-A^{\rm e}-B^{\rm e} r - D^{\rm e} r^2\,,\;\hfill r_0\leq r
<2\;$\cr}\eqno(12)$$
where the constants $A^{\rm e}$, $B^{\rm e}$ and $D^{\rm e}$ are given 
in Table 1.

The absolute error in the fit is less than $\simeq 10^{-3}$ above $r=r_0$. 
Below $r=r_0$ the error is in principle not kown. Yakovlev and Shalybkov (1989)
calculated the value of the classical contribution from polarized electrons
$C^{\rm e}_Z(\Gamma)$, using the so-called linear mixing 
law, whose accuracy has been demonstrated recently (Rosenfeld 1994, 1996;  DeWitt, Slattery \& Chabrier 1996). However, Yakovlev and Shalybkov's calculations are
based on an expansion scheme of the free energy due to the electron polarization
with respect to the reference rigid background energy, thus reducing the
validity of the calculations to the weak screening (small $x$ or $r_s$) regime.
Chabrier (1998) extended these calculations to regimes of stronger screening by
calculating explicitely the electron contribution from eqn(5). The equations
were solved within the framework of the so-called hyper-netted chain (HNC) theory of the N-body problem. The HNC results are found to agree remarkably well
($< 1\%$) with the Monte-Carlo results (see e.g. Chabrier 1990; DeWitt, Slattery \& Chabrier 1996). Given the rapidity of the HNC
calculations compared to lengthy numerical MC-simulations, we were able to
explore a wide density- and temperature-range corresponding to different
values of the relativistic parameter.
We 
used these calculations of $C^{\rm e}_Z(\Gamma)$ and the small-$r$ behaviour of $V^{\rm e}(\rho,r)$
to calculate $\beta w(0)=-h_0^{\rm e}=-C^{\rm e}_Z(\Gamma)-\lim_{r\rightarrow
 0}(\beta V^{\rm e}(\rho,r)-\Gamma/r)$.
The constants $h_1^{\rm e}$, $h_2^{\rm e}$ and 
$r_0$ are found by extrapolating the Monte Carlo results in a way
similar to Rosenfeld (1992) and Sahrling (1994b). The expressions for these constants are 
quite complicated and will not be given explicitly.

\subsection{Enhancement Factors}
We have calculated the enhancement factors due to the electron gas
polarization.
The correction with respect to the rigid-background calculations, for the
one-component case, is given by :
$$E^{\rm e}= {E\over E^{\rm OCP}}={\exp\left(C^{\rm e}_Z-{3\Gamma\over
\delta}~f_Z(\delta)
\right)}\; \eqno(13)$$

where $C^{\rm e}_Z(\Gamma)$ denotes the classical contribution from the
polarized electrons and the second term is the quantum correction. 
We use Eq. (10) to calculate $E$ for $Z=6$ and $8$. For the rigid background
(OCP) result we use a similar
expression but with the potentials derived in Rosenfeld (1992; 1994). For $C^{\rm e}_Z$ the results calculated at $x=10$ can be parametrized by :

$$C^{\rm e}_6(\Gamma)= 0.0123\, \Gamma + 0.0125\, \Gamma^{1/4}- 0.00554~,$$
$$ f_6(\delta) = 5.26~10^{-3} ~\delta - 8.76~10^{-4}~\delta^2 +  3.94~10^{-5}~\delta^3~,$$
$$ C^{\rm e}_8(\Gamma) = 0.0160\, \Gamma + 0.0350\, \Gamma^{1/4} - 0.0326~,$$
$$ f_8(\delta) =5.38~10^{-3} ~\delta - 4.51~10^{-4}~\delta^2 +  4.96~10^{-6}~\delta^3~
\eqno(14)$$

As mentioned above, the $x$-dependence of the screening potential, and thus of
$C_Z^e$ is weak over the density-range of interest.
We found that these expressions for $C^{\rm e}_Z$ are valid in the regime
$2 < x < 20$, $\Gamma \le 60$ with an absolute rms-error less than 0.05.

The functions $f_Z$ have been calculated using a root-mean-square fit to
30 points in the interval $0<\delta\leq 3$. There is actually a weak dependence
on $\Gamma$ in $f$ but $f$ in Eq. (14) has been averaged to give a total
error in $E^e$ $\la 20 \%$ for the density- and temperature-range explored
presently. 
In Table 2 we compare these new results for the contribution arising from 
polarized electrons with the ones obtained in Sahrling (1994b). Within the accuracy of the present formulae, the electrons will essentially
{\it increase} the reaction rate.
It is clear from
the table that the extrapolation error made in Sahrling (1994b) is most severe for the
charge $Z$ but also that a careful analysis of different densities and temperatures 
is important.
 It is
noteworthy that quantum effects (high $\delta$) act to reduce the enhanced
reaction rate caused by the lowering of the Coulomb barrier, as given by $C^{\rm e}_Z$. For
high values of $\delta$ and $\Gamma$, quantum effects can be very important for 
$Z=8$.
In all cases, the contribution of the electrons to the enhancement factors of
nuclear reactions in dense stellar plasmas remains of the order of the unity,
at most a factor $\sim 10$ if quantum corrections are not included. This is
much smaller than the classical {\it ionic} contribution $E^{OCP}\approx e^\Gamma$
under similar
thermodynamic conditions. Confusion
had been brought by Ichimaru and collaborators (Ichimaru \& Utsumi 1983; Ichimaru \& Ogata 1991) who obtained electron enhancement factors of several
orders of magnitude. These results are based on erroneous calculations which
involve the difference between two very large numbers (eqns (17) and (9) of
the afore-mentioned references, respectively), whereas a direct resolution
of eqn. (10) with the potential (5) can be done easily, as in the present calculations.

\section{Carbon Ignition curve}
In the late stages of stellar evolution for intermediate-mass stars, the core 
consists of a mixture of carbon and oxygen ions, and degenerate electrons. 
As
density and temperature increase, they will reach eventually thermodynamic
conditions  
where the energy release from the nuclear reactions 
equals locally
the neutrino energy loss. The points in the $\rho-T$ 
plane where this
occurs define the carbon ignition curve. This is of prime importance in particular for accreting white dwarfs, for it determines the fate of the object.
Small changes in the enhancement factors of nuclear reactions can make a
spectacular difference in the outcome, the star becoming either a type-I
supernova or collapsing into a neutron star (Isern \& Hernanz 1994). This
demonstrates the importance of accurate calculations for these factors.
With the enhancement factor $E=E^{OCP}E^e$ in Eq.
(13), the generalisation to mixtures according to Rosenfeld (1992; 1994), the 
perfect gas
reaction rates of Caughlan \& Fowler (1988) and the neutrino
rates calculated by
Itoh et al. (1989; 1992), one gets the curves 
shown in Fig.
2 for different oxygen abundances. Note that although the expressions for the 
electron screening differ by up to a factor 2 with respect to the results of Sahrling
(1994b), the corresponding change in the carbon ignition curve is only about 10 \% at a 
given temperature around $\Gamma=180$.

Note also that the effect of the electron screening on the carbon ignition curve
remains small, as expected from the value of $E_e$ close to unity, contrarily to what has been claimed by Ichimaru \& Ogata (1991).
These latter calculations were based on the afore-mentioned erroneous electron enhancement
factors. As shown on Figure 2, the full and dotted lines (pure carbon with and without polarization, respectively) are nearly superimposed.
As shown by Chabrier (1993), quantum effects in the structure of the ionic
stellar plasma might also affect the plasma melting curve, and thus the WD cooling and the ignition curve. Work in this
direction is under progress.  


\section{Conclusions}
We have calculated enhancement factors for non-resonant nuclear reaction
rates due to polarized electrons over a wide range of 
temperatures and densities characteristic of dense stellar plasmas. The
calculations have been conducted for large values of the electron relativistic
parameter and thus are not restricted to weak screening.
The analysis is based on a new set of mean (or screening-) 
potentials calculated using standard Monte Carlo techniques and N-body theory
calculations. We have focussed on carbon and oxygen 
ions since one expects high-order quantum effects to be most important for 
these nuclei. The calculations can be easily applied to similar mixtures
as oxygen-neon-magnesium for example. We find that quantum high-order
corrections to the classical enhancement factor will be important for low 
temperatures and high densities and will {\it reduce} the zeroth order
(classical)
contribution. In an earlier work we discussed the possibility that 
polarized electrons would reduce the nuclear reaction. This work was based
on a mean potential calculated for carbon only and just one density and 
temperature, so the extrapolation error was expected to be large. The 
present work has removed this extrapolation error and we have shown that 
polarized electrons yield a small ($\sim 1-2$) enhancement of the nuclear
reaction rate, contrarily to what had been claimed in previous studies. These results have been used to derive more accurate carbon
ignition curves for the ultimate stages of intermediate-mass stars and accreting
white dwarfs.

\bigskip 

{\bf Acknowledgments :} The authors wish to thank Christopher Pethick for many
interesting discussions on the physics of nuclear reaction rates. Bengt Gustafsson
 has also
been of substantial help in the reading of the manuscript.
This work has been supported by travel grants from NORDITA, Copenhagen.
\vfill \eject

\section*{References}

Alastuey, A., Jancovici, B.: 1978, ApJ 226, 1034

Brush S.G., Sahlin H.L., Teller E.: 1966, Journ. of Chemical Physics 45, 2102

Canal R., Schatzman E., 1976, A\&A 46, 229

Caughlan, G.R., Fowler, W.A.: 1988, Atomic \& Nuclear Data Tables 40, 283

Chabrier G.: 1990, J. de Physique 51, 1607

Chabrier G.: 1993, ApJ 414, 695

Chabrier G.: 1998, in preparation

DeWitt, H.E., Slattery, W. and Chabrier, G., 1996, Physica B, 228, 21

Dubin, D., 1990, Phys. rev. A, 42, 4972 

Feynman, R.P., and Hibbs, A.R.: 1965, Quantum mechanics and Path 
Integrals. McGraw-Hill, New York

Gould, R.J.: 1990 ApJ  363, 574

Galam, S. \& Hansen J.P., 1976, Phys. Rev. A, 14, 816

Hubbard, W. B. \& Slattery, W., 1971, ApJ 168, 131



Ichimaru, S., Utsumi, K.: 1983, ApJ  269, L51

Ichimaru, S., Ogata, S.: 1991, ApJ  374, 647

Itoh, N., Adachi, T., Nakagawa, M., Kohyama, Y.: 1989, ApJ  339, 354

Itoh, N., Mutoh, H., Hikita A., Kohyama, Y.: 1992, ApJ  395, 622

Isern, J. \& Hernanz, M., 1994, IAU Colloquium \#147, {\it Equation of state in Astrophysics},  Chabrier G \& Schatzman E. Eds., Cambridge University Press. 

Jancovici, B., 1962, Nuovo Cim. 25, 428

Jancovici, B., 1977, J. Stat. Phys., 17, 357

J\"andel M., Sahrling M.: 1992, ApJ  393, 679

Mochkovitch R., Hernanz M., 1986, in {\it Nucleosynthesis and Its Implications on 
Nuclear and Particle Physics}, Audouze J. and Mathieu N. (eds.). Reidel,
Dordrecht, p. 407

Rosenfeld, Y.: 1992, Phys. Rev. A 46, 1059 

Rosenfeld, Y.: 1996, Phys. Rev. E 54
   
Rosenfeld Y.: 1994, IAU Colloquium \#147, {\it Equation of state in Astrophysics},  Chabrier G \& Schatzman E. Eds., Cambridge University Press.

Sahrling, M.: 1994 a,   A\&A 283, 1004 
 
Sahrling, M.: 1994 b,   A\&A 284, 484
  
Schatzman, E.: 1948, J. Phys. Rad. 9, 46

Slattery W.L., Doolen G.D. and DeWitt H.E.: 1980, Phys. Rev. A 21, 
2087


Yakovlev, D.G., Shalybkov, D.A.: 1989  Sov. Sci. Rev. 4 , part 7

\vfill\eject


\begin{table}[htbpe]
\caption{The constants in the mean potential Eq.(12)}
\bigskip
\begin{tabular}{llll}\hline
$Z$&$A^{\rm e}$&$B^{\rm e}$&$D^{\rm e}$
\\ \hline
6 & 1.055 & -0.541 & 0.071 \\
8 & 1.053 & -0.566 & 0.080   \\ \hline
\end{tabular}
\end{table}

\vskip 2.cm 

\begin{table}[htbpe]
\caption{Comparison with previous estimates of $E^{\rm e}$. The last column shows
the {\it classical} contribution to the enhancement factor calculated by 
Chabrier (1998). The numerical error in the calculation of $C^e$ is of the order
of 10 to 15\% for $\Gamma \wig > 70$.}
\bigskip

\begin{tabular}{lllllll}\hline
$Z$&$\Gamma$&$\delta$&x&$\rm Sahrling (1994b)$&$E^{\rm e}$&$\exp(C^{\rm e}_Z)$\\ \hline

8 & 113.8 & 0.8 & 5.0 & 0.62 & 1.2 & 6.69 \\
8 & 36.4 & 0.37 & 5.0 & 1.2 & 1.1 & 1.89 \\
8 & 72.6 & 0.74 & 10.0 & 0.82 & 1.14 & 3.42\\
8 & 145.2 & 1.18 & 10.0 & $0.38$ & 1.34 & 11.1\\
8 & 46.0 & 0.69 & 20.0 & 0.93 & 1.1 & 2.21 \\ \hline
6 & 70.4 & 0.8 & 5.0 & 0.75 & 0.93 & 2.45  \\
6 & 22.5 &  0.37 & 5.0 & 1.1 & 0.97 & 1.35 \\
 6 & 44.9 & 0.74 & 10.0 & 0.89 & 0.96  & 1.78\\
 6 & 89.9 & 1.18 & 10.0 & $0.56$ & 0.98 & 3.12\\
6 & 142.9 & 2 & 20.0 & 0.38 & 1.3 & 5.99\\
6 & 28.5 & 0.69 & 20.0 & 0.96 & 0.97 & 1.45 \\
\end{tabular}
\end{table}
\vfill \eject

\vfill \eject \onecolumn

\centerline {\bf FIGURE CAPTIONS} \vskip1cm

\indent{\bf Figure 1 :} Screening potential for $\Gamma=120, x=10, Z=6$. The solid line 
corresponds to the fitting formula in Eq. (12). The dots indicate the Monte Carlo 
result. The systematic error in the Monte Carlo dots is $\simeq 10^{-4}$ 
unless explicitly shown by vertical bars

\vskip1cm

\indent{\bf Figure 1b :} Same as a) but with $Z=8$

\vskip1cm

\indent{\bf Figure 2 :} The carbon ignition curve for pure carbon and a 50\%
carbon-oxygen mixture. For comparison we also show the curves obtained 
for pure carbon 
by Ichimaru and Ogata (1991) (IO91), Sahrling (1994b), and within the 
OCP (rigid background) approximation.
\vfill \eject

\begin{figure}
\epsfbox{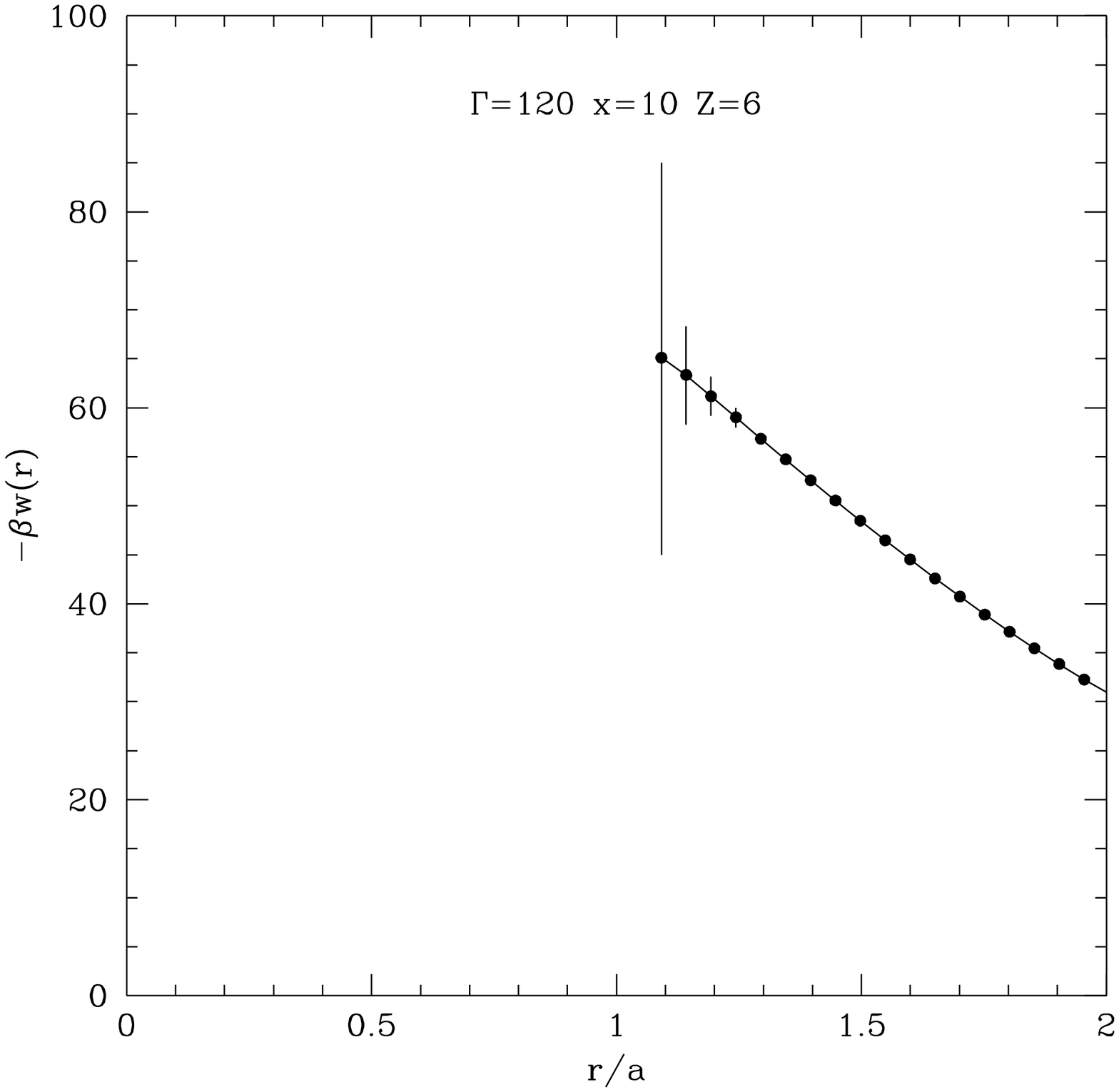} 
\end{figure}

\begin{figure}
\epsfbox{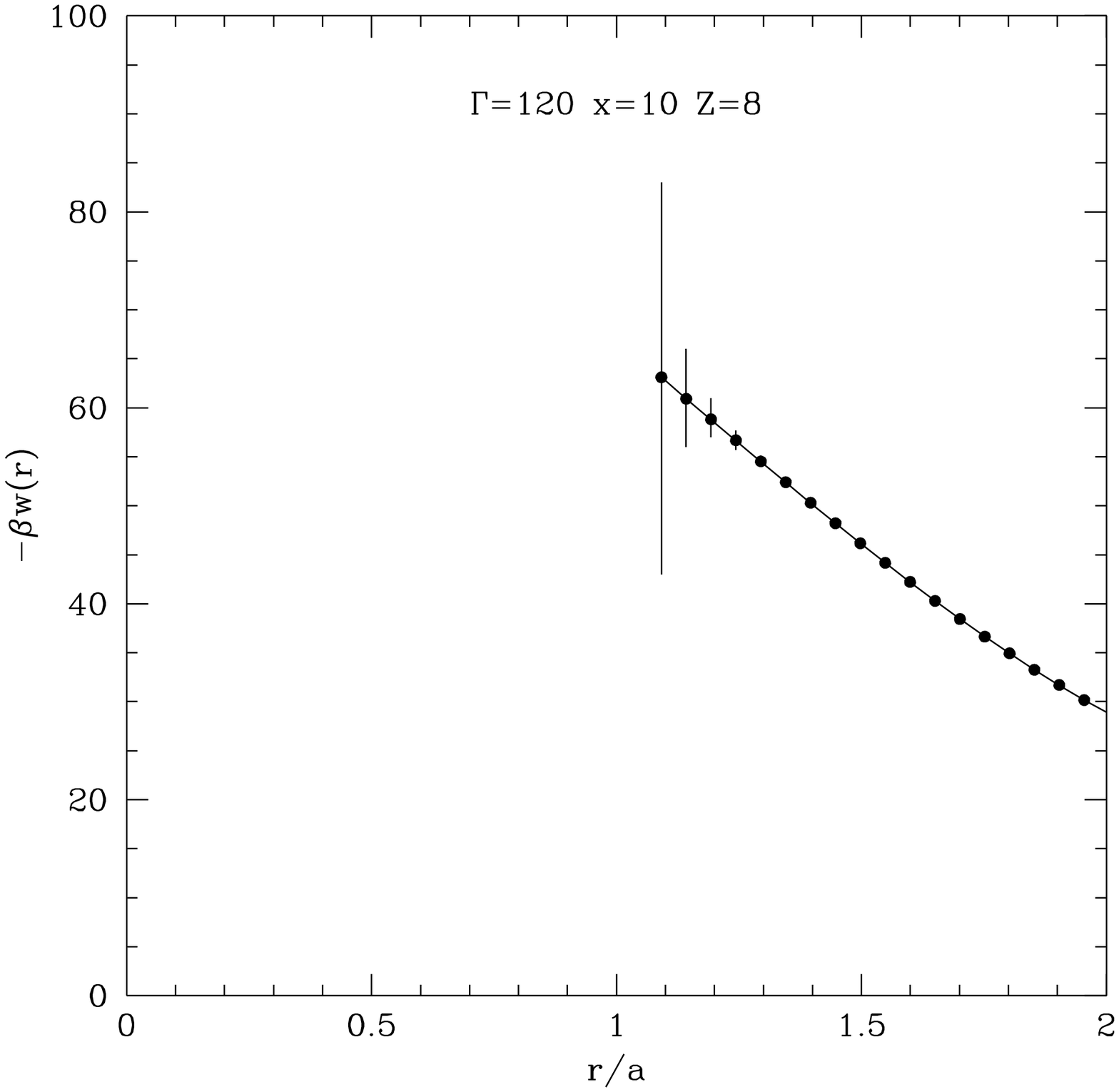} 
\end{figure}

\begin{figure}
\epsfbox{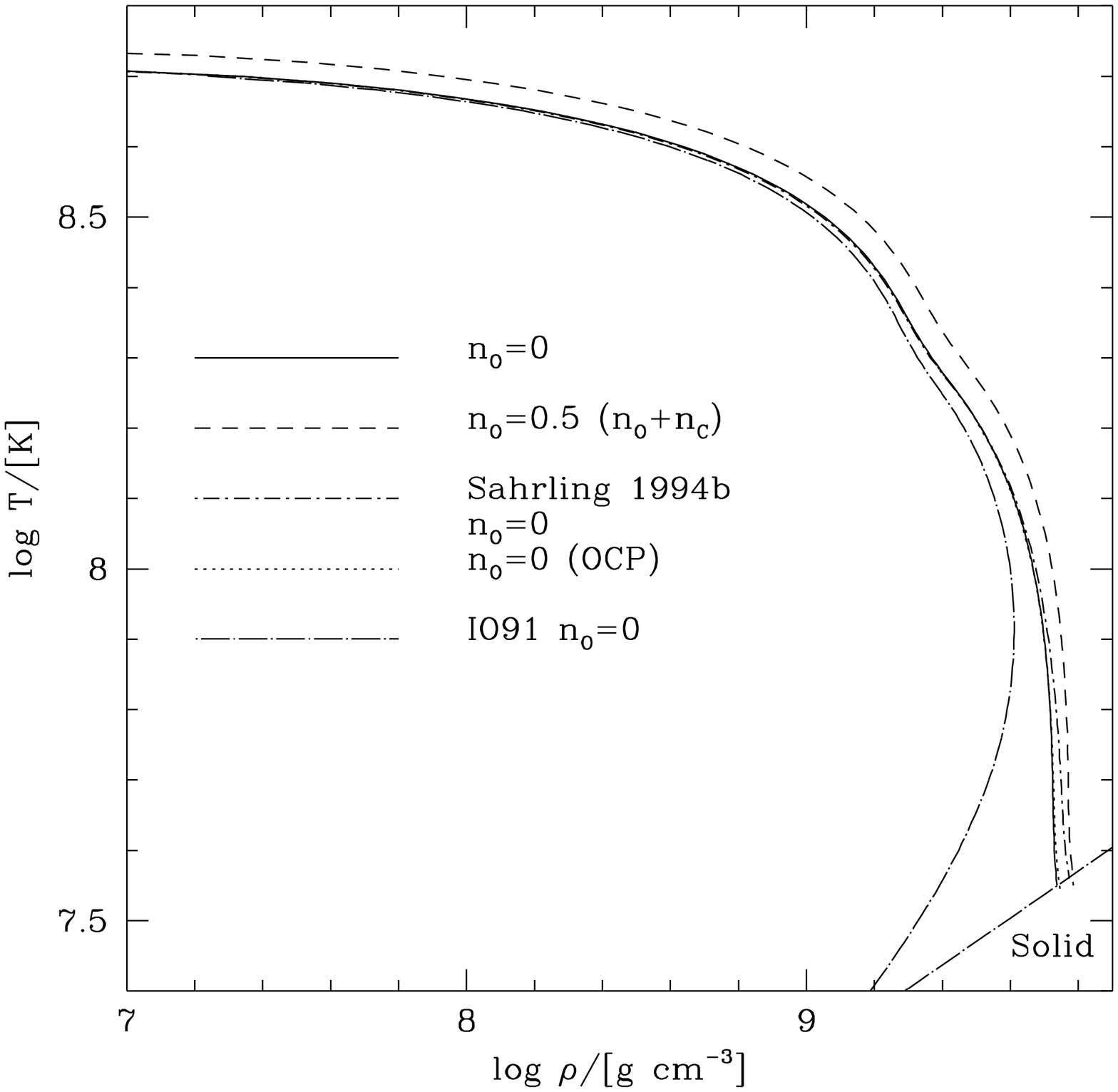} 
\end{figure}

\end{document}